\documentclass[aps,prl,twocolumn,showpacs,floatfix]{revtex4}
\usepackage{graphicx}
\usepackage{dcolumn}
\usepackage{bm}
\usepackage{xspace}
\usepackage{pstricks}

\newcommand{\cross}{\mathsf{X}}
\newcommand{\Eref}[1]{Eq.~(\ref{eq:#1})}
\newcommand{\eref}[1]{(\ref{eq:#1})}
\newcommand{\elabel}[1]{\label{eq:#1}}
\newcommand{\Fref}[1]{Fig.~\ref{fig:#1}}
\newcommand{\flabel}[1]{\label{fig:#1}}
\newcommand{\etal}{{\it et~al.\@\xspace}}
\newcommand{\ave}[1]{{\langle #1 \rangle}}
\newcommand{\suboOM}{o}
\newcommand{\subaOM}{a}
\newcommand{\oOM}{oOM\xspace}
\newcommand{\aOM}{aOM\xspace}
\newcommand{\ie}{i.e.\xspace}
\newcommand{\GC}{{\mathcal G}}

\newcommand{\PCs}{{\mathcal P}(s;L,p_r,p_l)}
\newcommand{\rough}{{\chi}}
\newcommand{\prepsection}[1]{}

\begin{document}
\title{Anisotropy and universality: the Oslo model, the rice pile experiment
and the quenched Edwards-Wilkinson equation.}
\author{Gunnar Pruessner}
\email{gunnar.pruessner@physics.org} 
\homepage{http://www.ma.ic.ac.uk/~pruess/}
\author{Henrik Jeldtoft Jensen}
\email{h.jensen@ic.ac.uk} 
\homepage{http://www.ma.ic.ac.uk/~hjjens/}
\affiliation{Department of Mathematics, Imperial College, 
180 Queen's Gate, London SW7 2BZ, U.K.}
\date{16.07.03}

\begin{abstract}
We show that any amount of anisotropy moves the Oslo model to another
 \emph{known} universality class, the exponents of which can be derived
 exactly. This amounts to an exact solution of the quenched
 Edwards-Wilkinson equation with a drift term.  We argue that anisotropy
 is likely to be experimentally relevant and may explain why consistent
 exponents have not been extracted in the rice pile experiments.
\end{abstract}

\maketitle

\prepsection{Introduction} The suggestion that power-law like
distributions frequently reported in experiments may be the effect of
threshold dynamics and metastability was made by Bak, Tang and
Wiesenfeld in 1987 \cite{BakTangWiesenfeld:1987} and was called Self-Organized Criticality (SOC). Avalanche
dynamics in granular piles was from the onset used as a metaphor and
laboratory for SOC behavior and has inspired many models and experiments
\cite{Jensen:98}. One of the most celebrated of these efforts is
the experimental study of avalanches in one dimensional rice piles by
Frette \etal \cite{FretteETAL:1996} and the theoretical Oslo model
\cite{ChristensenETAL:1996} inspired by the rice pile experiment. The
general interest and relevance of such studies rely on the assumption,
guided by equilibrium critical phenomena, that the critical behavior of
scale invariant systems falls into universality classes determined solely
by a few general characteristics of the system, such as symmetry and
dimension. So-called ``relevant'' parameters can decide which of the
symmetries the system is asymptotically dominated by.

The r\^{o}le of anisotropy in SOC has been highlighted very early by Hwa
and Kardar \cite{Hwa:1989} and Grinstein \etal~\cite{Grinstein:1990},
who used anisotropic Langevin equations to describe sandpiles. On the
cellular automata level, Kadanoff \etal~\cite{KadanoffETAL:1989} have
conjectured that the net flux of particles is a relevant parameter. In
this paper we confirm this conjecture for the Oslo model, which shows a
clear cut and consistent relevant dependence on anisotropy. This is of
great importance for the interpretation of experimental results
\cite{FretteETAL:1996} and more generally for the much studied quenched
Edwards-Wilkinson equation \cite{NattermannETAL:1992,Pruessner:2003}.

Moreover, contrary to suggestions in former studies, the switch between
different universality classes (crossover) is \emph{not} triggered by
the introduction of stochasticity
\cite{DharRamaswamy:1989,TadicDhar:1997,Pastor-SatorrasVespignani:2000a,KlosterMaslovTang:2001}
nor by multiple topplings
\cite{Vazquez:2000,Pastor-SatorrasVespignani:2000a,PaczuskiBassler:2000,PriezzhevETAL:2001}.

Similar to \cite{TsuchiyaKatori:1999}, the system size at which the
crossover occurs, $L_\cross$, depends on the strength of the anisotropy
$v$. We exemplify two possible mechanisms causing anisotropy in
experiments, one of which vanishes with system size $L$ fast enough to
keep $v L$ constant. This represents a marginal case and consequently
makes a unique identification of the critical exponents impossible.

\prepsection{The model} The original Oslo model (\oOM) consists, in one
dimension, of a lattice of sites $i=1,\dots,L$. Two coupled dynamical
variables are associated with each lattice site: the primary variable
$z_i\in \{0,1,2,\dots\}$ and the threshold variable
$z^c_i\in\{1,2\}$. The initial configuration consists of $z_i=0 \;
\forall i$ and a random configuration of the $z^c_i$. The system is
driven by increasing $z_1$ by one (a ``grain'') followed by a relaxation
of all sites $1\leq i\leq L$ for which $z_i>z_i^c$ (``over-critical''
sites). In case a site $i$ is over-critical, the following updates are
performed (``toppling'' or ``relaxation''): $z_i\rightarrow z_i-2$ and
$z_{i\pm1}\rightarrow z_{i\pm1}+1$ and, importantly, the existing value
of the threshold $z_i^c$ is afterwards replaced by 1 with probability
$p$ and by 2 with probability $1-p$.  The boundaries are updated the
same way except that for $i=1$ ($i=L$) addition on site $0$ ($L+1$) is
omitted.

We now introduce a tunable degree of anisotropy into the dynamics. An
over-critical site $i$ is relaxed in the following way. Only left
movement: with probability $p_l(1-p_r)$ perform the updates:
$z_i\rightarrow z_i-1$ and $z_{i-1}\rightarrow z_{i-1}+1$. Only right
movement: with probability $p_r(1-p_l)$ perform the updates
$z_i\rightarrow z_i-1$ and $z_{i+1}\rightarrow z_{i+1}+1$.  Both left
and right movement: with probability $p_l p_r$ perform the updates
$z_{i\pm1}\rightarrow z_{i\pm1}+1$ and $z_i\rightarrow z_i-1$. A new
$z^c_i$ is chosen, at random, after every successful update, \ie when
at least one grain has been redistributed. We call this version of
the model the anisotropic Oslo model (\aOM). The strength of the
anisotropy is described by the drift velocity $v=(p_r-p_l)/(p_r+p_l)$
which is the net flux of grains through the system. Clearly it is only
sensible to study $p_r+p_l>0$. The case $p_r=p_l=1$ corresponds exactly
to the \oOM, while $p_r=p_l\ne 1$ represents a stochastic variant of the
\oOM. The avalanche exponents for the extreme, totally asymmetric case
$p_l=0$ and $p_r=1$ can be obtained exactly
\cite{Pruessner_exactTAOM:2003} and describe, as we shall see below, the
scaling behavior for all $v>0$. We are interested in the statistics of
the sizes, $s$, of the avalanches of relaxation induced by the driving
$z_1\rightarrow z_1+1$. The size of an avalanche is in both versions of
the model defined as the number of times the relaxation rule was
successfully applied after the drive
$z_1\rightarrow z_1+1$ in order to make $z_i<z_i^c\; \forall_i $ yet
again. Thus $s\geq 0$. From the definition above it is
clear that the model is equally well defined in terms of reduced
probabilities, omitting the case where no grain is redistributed.

In the totally anisotropic or asymmetric limit \cite{PriezzhevETAL:2001}
the model resembles some features of other exactly solved,
\emph{directed} models
\cite{DharRamaswamy:1989,MaslovZhang:1995,PaczuskiBassler:2000,KlosterMaslovTang:2001,PriezzhevETAL:2001}.
In two dimensions a very similar model has been studied numerically
\cite{Vazquez:2000}.  However, we stress that contrary to some other
``exact'' solutions, the model is solvable directly on the lattice and
without assuming any scaling behavior \cite{Pruessner_exactTAOM:2003}. Also
the amplitudes of the moments can be calculated exactly.

It is a tedious, but straightforward task to show that the \aOM is
``Abelian'', \ie the order of updates is irrelevant for its statistical
properties. Since the microdynamics which prescribes the order of
updates is irrelevant, there is no unique way to define a microscopic
timescale. Presumably universal exponents of the duration of avalanches
are therefore mainly a property of the arbitrary choice of the
microdynamics. According to Hughes and Paczuski
\cite{HughesPaczuski:2002}, a non-Abelian variant of an Abelian 
model may or may
not remain in the same universality class.

Moreover, one notes that the \oOM as well as the \aOM contains multiple
topplings, \ie a single site can relax several times during a single
avalanche.

\begin{figure}
\includegraphics[width=0.9\linewidth]{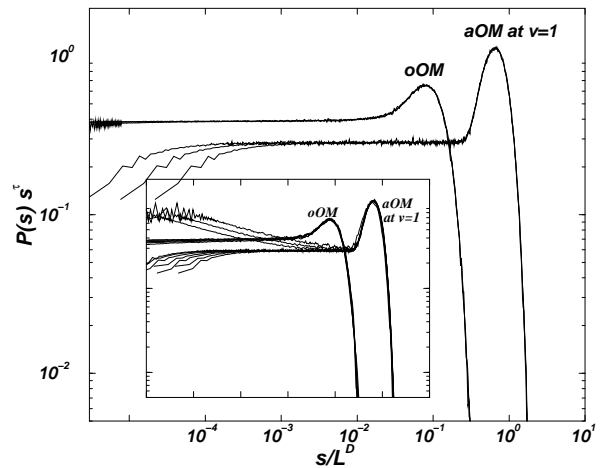}
\caption{\flabel{collapse_comparison} Main panel: Data collapse ($\PCs
s^\tau$ vs. $s/L^D$) of the normalized and binned $\PCs$ for the two extreme cases ($p_r=1$ with $p_l=0$ and
$p_l=1$) for $L=1280,2560,5120$. The PDF $\PCs$ has been rescaled using
$\tau=1.333\dots$ and $D=1.5$ for the \aOM and $\tau=1.555\dots$ and
$D=2.25$ for the \oOM. Inset: Rescaled PDFs for three different choices
of $p_r>p_l$, namely $(p_r,p_l)=(1.0,0.95), (1.0,0.25), (0.75, 0.25)$,
and $L=640, 1280, 2560$, and two choices of $p_r=p_l$, namely $0.75$ and
$0.25$ and the same range of $L$. The rescaled PDFs of each tuple
$(p_r,p_l)$ would form a single line, which fixes the two exponents
$\tau$ and $D$. By tuning the parameters $a$ and $b$ in
\Eref{simple_scaling}, as done in the inset, the resulting single line
collapses with one of the extreme cases plotted in the main panel and
the inset. Deviations for intermediate values of $s/L^D$ (\ie away from
the bump towards smaller values of $s$) are expected to vanish at
sufficiently large $L$.  }
\end{figure}
\begin{figure}
\includegraphics[width=0.9\linewidth]{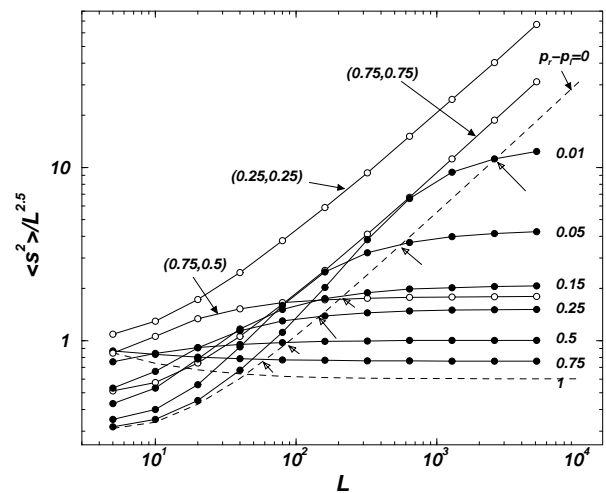}
\caption{\flabel{crossover} 
Scaling of $\ave{s^2}/L^{2.5}$ for different anisotropies. The filled
 circles are results for $p_r-p_l$ as indicated and $p_r=1$. Open circles show
 other parameters $(p_r,p_l)$. The dashed lines are the two extreme
 cases \oOM ($p_r=p_l=1$) and the solvable model ($p_r=1, p_l=0$). The open
 arrows mark the approximate crossover.} 
\end{figure}
\begin{figure}
\includegraphics[width=0.9\linewidth]{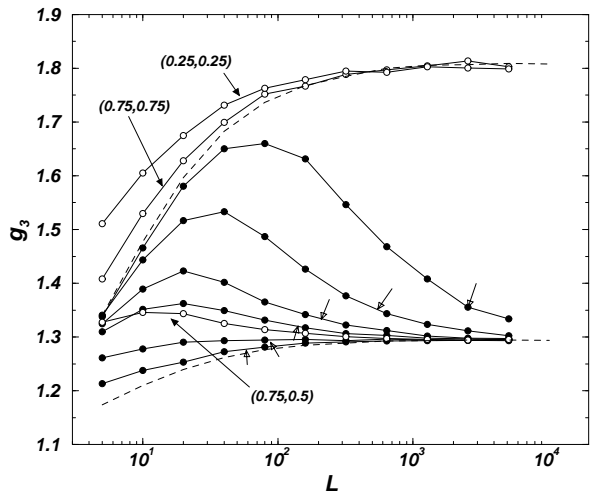} \caption{\flabel{g3}
Scaling of $g_3$ (see \Eref{gn}) for different anisotropies. The filled
circles are results for $p_r-p_l$ in the same vertical order as shown in
\Fref{crossover} and $p_r=1$. Open circles show other parameters
$(p_r,p_l)$ as indicated. The dashed lines are the two extreme cases \oOM
($p_r=p_l=1$) and the solvable model ($p_r=1, p_l=0$). 
The open arrows mark the crossover points in \Fref{crossover}.}
\end{figure}

\prepsection{Results} 
We now describe the avalanche statistics of the \aOM.  In
\Fref{collapse_comparison} we demonstrate that the avalanche size
distribution $\PCs$ follows simple (finite size) scaling
\begin{equation} \elabel{simple_scaling}
 \PCs = a(p_r,p_l) s^{-\tau}\GC \left(\frac{s}{b(p_r,p_l) L^D}\right) \text{ for } s>s_l
\end{equation}
where $\GC$ is the universal scaling function, $a(p_r,p_l)$ and
$b(p_r,p_l)$ are two anisotropy and system dependent parameters, and
$s_l$ is the lower cutoff independent of $L$. The values for the scaling
exponents are $\tau_{\subaOM}=4/3$ and $D_{\subaOM}=3/2$, which can be
derived exactly in the asymmetric limit $v=1$
\cite{Pruessner_exactTAOM:2003} and represent the known universality
class of the stochastic, directed sandpile in \emph{two} dimensions
\cite{Pastor-SatorrasVespignani:2000a,PaczuskiBassler:2000,KlosterMaslovTang:2001},
which is in turn closely related to the directed sandpile
\cite{DharRamaswamy:1989}. Numerically, these
exponents have been found for all $v>0$ studied at sufficiently large
system sizes $L \gg L_\cross$. The crossover that occurs around
$L_\cross$ is discussed in detail below. The two exponents
of the \aOM are to be compared with the
exponents for the \oOM, of $\tau_{\suboOM}=1.556(4)$ and
$D_{\suboOM}=2.25(2)$. Since the average avalanche size scales linearly
with the system size, the exponents are related by $D(2-\tau)=1$.
\cite{ChristensenETAL:1996,PaczuskiBoettcher:1996}

The easiest way to derive the exponents from numerical data is by
analysis of the moments \cite{TebaldiDeMenechStella:1999}, which scale
according to \eref{simple_scaling} for $n>\tau-1$ in leading order like
\begin{equation} \elabel{moments}
 \ave{s^n} = \int_0^\infty ds s^n\PCs = a (b L^D)^{1+n-\tau} g_n + \dots
\end{equation}
where $g_n$ is discussed below and $\dots$ denotes subleading terms, especially Wegner's corrections
to scaling \cite{Wegner:72}. In the following, the crossover is studied
by means of the rescaled second moment, $\ave{s^2}/L^{5/2}$, which is
shown in \Fref{crossover}. For non-vanishing anisotropy, $v>0$, it
approaches a constant as $L\to\infty$. For very small but finite values
of $v$ and $L$ the rescaled moment increases with $L$ like $L^{0.75}$,
corresponding to the \oOM behavior, but at $L \approx L_\cross(v)$ it
crosses over and eventually converges to a finite constant.  Below, we
shall relate the behavior of $L_\cross(v)$ to the effective anisotropy
relevant to an experiment of a given size. Here we emphasize that
\Fref{crossover} clearly demonstrates that the universality class of the
extremely anisotropic case, $p_r=1$, $p_l=0$, contains all systems with
non-vanishing anisotropy $v>0$. That renders the \oOM with $v=0$ a
special case; however, it is remarkable that even for $p_r=p_l\ne 1$ the
model still shows \oOM behavior. Thus, it is not the stochasticity
itself
\cite{DharRamaswamy:1989,TadicDhar:1997,Pastor-SatorrasVespignani:2000a}
which induces the change in critical behavior.

\prepsection{Universal Amplitude Ratios}
\Eref{simple_scaling} allows the definition of universal amplitude ratios
\begin{equation} \elabel{gn}
 g_n = \frac{\ave{s^n} \ave{s}^{n-2}}{\ave{s^2}^{(n-1)}}
\end{equation}
which can easily be proven to be asymptotically independent of $a$, $b$
and $L$. The two constraints on $\GC$, which fix $a$ and $b$ in
\eref{simple_scaling}, can be chosen such that $g_n$ are the moments of
$x^{-\tau}\GC(x)$ as used in \eref{moments}, namely by imposing that $\int_0^\infty
x^{1-\tau}\GC(x) = \int_0^\infty
x^{2-\tau}\GC(x)=1$. The universal
amplitude ratio $g_3$ as shown in \Fref{g3} indicates not only the same
crossover behavior as observed in \Fref{crossover}, but also the
universality of $\GC$.

The importance of the above result is highlighted when we recall
that the \oOM in the continuum limit is described by the quenched
Edwards-Wilkinson (EW) equation \cite{Pruessner:2003}. A similar
derivation shows that the \aOM is a quenched EW equation with
an additional drift term 
\begin{equation} \elabel{EW}
\partial_t h(x,t) = D \partial_x^2 h(x,t) - v \partial_x h(x,t)  +\eta(x,h(x,t)),
\end{equation}
where $D$ is the surface tension and $v$ the anisotropy or drift
velocity as defined above. In the \aOM, the quenched
noise $\eta(x,h(x,t))$ represents the randomly chosen $z^c_i$ and
$h(x,t)$ is the number of charges received by site $x$ at time $t$. The
quenched nature of the noise makes it difficult to solve \eref{EW}
directly. Together with the
boundary conditions \cite{Pruessner:2003}, it prevents the drift term $v
\partial_x h$ from being absorbed by a Galilean transformation. However, the
above results determines the roughness exponent via $D=1+\rough$
\cite{PaczuskiBoettcher:1996} to be $\rough =1/2$ for $v>0$, which has
already been suggested in another case of anisotropic depinning
\cite{TangKardarDhar:1995}. For $v=0$ numerics for the \oOM suggest
correspondingly that
$\rough=1.25(2)$.

The crossover for small $v>0$ with increasing $L$ can also be
illuminated by a study of the individual grains in the system, which
behave like biased random walkers. This leads again to a diffusion
equation with drift term and two absorbing boundaries
\cite{FarkasFueloep:2001}. The average avalanche size is the average
time the particles spend in the system divided by the average number of
grains redistributed per toppling. The crossover is expected as soon as
the ballistic motion dominates over the diffusion, $L^2/D > L/v$, thus
$L_\cross(v)=D/v$. This has also been tested numerically, based on
heuristic estimation of $L_\cross$, as shown by the marks in
\Fref{crossover}.

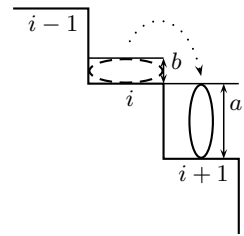
\begin{figure}[th]
\begin{center}
\begin{pspicture}(0,0)(4.0,3.2)
\put(0,-1.3){
\psline%
(0.0,4.0)%
(1.0,4.0)%
(1.0,3.0)%
(2.0,3.0)%
(2.0,2.0)%
(3.0,2.0)%
(3.0,1.0)
\psellipse(2.5,2.5)(0.17,0.5)
\psellipse[linestyle=dashed](1.5,3.17)(0.5,0.17)
\psbezier[linestyle=dotted]{->}(1.5,3.45)(1.7,4.0)(2.3,4.0)(2.5,3.1)
\psline[linewidth=0.2mm](2.0,3.0)(3.0,3.0)
\psline[linewidth=0.2mm]{<->}(2.8,2.0)(2.8,3.0)
\put(2.88,2.65){$a$}

\psline[linewidth=0.2mm](1.0,3.34)(2.0,3.34)
\psline[linewidth=0.2mm]{<->}(2.0,3.0)(2.0,3.34)
\put(2.1,3.2){$b$}

\put(1.5,2.7){$i$}
\put(0.2,3.7){$i-1$}
\put(2.2,1.7){$i+1$}
}
\end{pspicture}
\end{center}
\caption{\flabel{rice_toppling}
A stylized toppling of a single grain: If an elongated grain of width
$b$ and height $a$ topples from site $i$ to site $i+1$, it reduces the
height at $i$ by $b$ and increases the height at $i+1$ by $a$, thereby
increasing the slope at site $i-1$ by $b$, decreasing the slope at $i$
by $a+b$ and increasing the slope at $i+1$ by $a$. The net flux of slope
is $a-b$ to the right.  }
\end{figure}

\prepsection{Experiment} We now discuss the relevance of anisotropy to
real experimental granular systems. We stress that the anisotropy is in
the amount of {\em slope} transported between sites involved in a
relaxation event. Any net flux of the slope is eventually compensated by
the toppling of the last site and the slope is therefore asymptotically
stationary.

One process leading to an anisotropic
redistribution of slope arises when the toppling grain is elongated, see
\Fref{rice_toppling}. Most remarkably, in the original experiment
\cite{FretteETAL:1996} it was noted that only the elongated rice samples
showed scale invariant behavior. A reorientation of a single grain, as shown in \Fref{rice_toppling},
leads to a net flux of slope to the right. It can happen only once, and
in fact it depends on how and whether the rice enters and/or leaves the system
with a typical orientation. If that is the case, then an average
reorientation is distributed among all topplings on its way through the
system, \ie $v\propto 1/L$. Because $L_\cross
\propto 1/v$ and $v\propto 1/L$, this represents a marginal case,
because it is impossible to decide whether $L \gg L_\cross(v)$ or not.
The dependence of, say, the ratio $\langle s^2\rangle/L^{2.5}$ on $L$
would in this case be given by an (unknown) trajectory through the
diagram in \Fref{crossover} since a change in $L$ also leads to a change
in $v$, which allows non-universal quantities to
enter. 


We do not know if the experiment by Frette \etal \cite{FretteETAL:1996}
involves this complication, but the exponents extracted from the
experiment by the authors are not consistent\footnote{There are some
problems in the derivation: The scaling ansatz is unphysical, as higher
moments diverge even in finite systems. The result $\beta=\nu$ is due to
a neglect of the lower cutoff. Moreover, the possible interpretation of
$\alpha\approx 2.02$ as $\tau$ leads to system size independent finite
first moment, which contradicts $\ave{s}\propto L$. In this light it
would be very interesting to repeat the data analysis.}.

Local rearrangements, such as expansions (on the site that looses a
grain) and compressions (on the site that receives the grain) lead to
an anisotropy which does not vanish with $L$: Say column $i-1$ from
which the grain leaves expands by $\epsilon(h_{i})$ and column $i+1$
which receives the grain is compressed by an amount $\epsilon(h_{i+1})$.
Then the changes of the slopes during toppling are $\Delta z_{i-1} =
1-\epsilon(h_{i})$, $\Delta z_i = -2 + \epsilon(h_i)+\epsilon(h_{i+1})$
and $\Delta z_{i+1}=1 - \epsilon(h_{i+1})$. Assuming that the columns
behave elastically, $\epsilon$ would be an increasing function of $h$
which results in a net flow to the right, \ie $v>0$. However, it
remains unclear whether any of these effects can be seen in
experimental systems. Moreover, having shown that anisotropy is a
relevant field, it would not be surprising to find other relevant
fields which lead to yet another universality class.

\prepsection{Summary and Conclusion} We have demonstrated that for any
amount of anisotropy the exponents of the original Oslo model change and
are given by simple rational numbers which can all be obtained exactly
\cite{Pruessner_exactTAOM:2003}. The crossover has been studied
numerically using a moment analysis, \Eref{moments}, and universal
amplitude ratios, \Eref{gn}, and the crossover length has been
determined. The generalized model described above continuously connects
the established original Oslo model and an exactly solvable, directed
variant. This variant has, compared to the original Oslo model, an
enormous basin of attraction, so that the latter may be regarded a
special case of the former.

Moreover, we find a change in critical behavior of an SOC model,
genuinely due to anisotropy, rather than stochasticity or the presence
of multiple topplings.

The results are theoretically interesting especially because of their
relation to the Edwards-Wilkinson equation, the roughness exponent of
which has been obtained in case of the presence of a drift term
in one dimension to be $\rough=1/2$. Moreover,
according to our study, experiments are seriously complicated due to a
coupling between system size and effective anisotropy. That might
provide a clue as to the apparent difficulties to find theoretically
predicted exponents in the real world.

\begin{acknowledgments}
It is a pleasure to thank Kim Christensen for very helpful discussions
 and Paul Anderson for proofreading. The authors gratefully acknowledge
 the support of EPSRC. G.P. would like to thank Quincy Thoeren for
 hospitality.
\end{acknowledgments}

\bibliography{articles,books}

\end{document}